\documentclass{PoS}
\usepackage{graphicx}
\usepackage{amsmath,amssymb,latexsym}
\usepackage{url}
\usepackage{bm}
\usepackage{xspace}
\usepackage{nicefrac}

\newcommand{\beq}{\begin{equation}}
\newcommand{\eeq}{\end{equation}}

\newcommand{\be}{\begin{equation}}
\newcommand{\ee}{\end{equation}}

\newcommand{\apj}{"Astrophys. J."}

\title{The Galactic Magnetic Field and UHECR Optics}

\ShortTitle{GMF and UHECRs}

\author{\speaker{Glennys R. Farrar$^{a}$}, Nafiun Awal$^{a}$,  Deepak Khurana$^{a}$ and Michael Sutherland$^{b}$\\
 a) Center for Cosmology and Particle Physics,  New York University, NY, NY \\
  b) Center for Cosmology and AstroParticle Physics, The Ohio State University, Columbus, OH \\
        E-mail: \email{gf25@nyu.edu}}
        
\abstract{A good model of the Galactic magnetic field is crucial for estimating the Galactic contribution in dark matter and CMB-cosmology studies, determining the sources of UHECRs, and also modeling the transport of Galactic CRs since the halo field provides an important escape route for by diffusion along its field lines. We briefly review the observational foundations of the Jansson-Farrar 2012 model for the large scale structure of the GMF, underscoring the robust evidence for a N-to-S directed, spiraling halo field.  New results on the lensing effect of the GMF on UHECRs are presented, displaying multiple images and dramatic magnification and demagnification that varies with source direction and CR rigidity.}  

\FullConference{The 34th International Cosmic Ray Conference,\\
		30 July- 6 August, 2015\\
		The Hague, The Netherlands}

\def\EeV{\ifmmode {\mathrm{Ee\kern -0.07em V}}\else
                   \textrm{Ee\kern -0.07em V}\fi\xspace}
\def\GeV{\ifmmode {\mathrm{Ge\kern -0.07em V}}\else
                   \textrm{Ge\kern -0.07em V}\fi\xspace}
\def\TeV{\ifmmode {\mathrm{Te\kern -0.07em V}}\else
                   \textrm{Te\kern -0.07em V}\fi\xspace}
\def\eV{\ifmmode {\mathrm{e\kern -0.07em V}}\else
                   \textrm{e\kern -0.07em V}\fi\xspace}
\def\meV{\ifmmode {\mathrm{me\kern -0.07em V}}\else
                   \textrm{me\kern -0.07em V}\fi\xspace}



\begin{document}

\section{The Galactic Magnetic Field}

Our understanding of the Galactic magnetic field (GMF) has improved tremendously in recent years. The Jansson-Farrar (2012) (JF12) GMF model is currently the most realistic and comprehensive model available \cite{jf12a,jf12b}.  In addition to the coherent component, the JF12 model describes the spatial variation of the random field strength and additionally has a striated random component, in both disk and halo.  ``Striated'' is the term coined by JF12 to describe a field component that is locally aligned along some direction but with a randomly changing sign, as would originate from a random field being stretched or compressed in some preferred direction, e.g., by a wind emanating from the Galactic plane.

The coherent and striated components of the JF12 model were constrained by fitting Faraday Rotation Measures of $\approx$ 40k all-sky extragalactic sources, simultaneously with WMAP polarized (Q,U)  synchrotron emission maps \cite{jf12a}.  They are proportional to the line-of-sight integrals of $n_{\rm e}B_{||}$, for the RMs and of $n_{\rm cre} B_{\perp}^{2}$, for Q \& U, where $n_{\rm e}$ and $n_{\rm cre}$ are the thermal and cosmic ray electron densities.  Fitting the WMAP total synchrotron emission map, allows the random field to be constrained as well \cite{jf12b}.  Altogether, there are more than 10,000 independent datapoints consisting of the mean RM, Q, U and I in 13.4 degree-squared Healpix pixels, each with its own astrophysical variance \emph{measured} from the variance among the16 sub-pixel measurements.  The variance comes from Galactic random fields, local inhomogeneities in $n_{\rm e, \, cre}$, and in the case of RMs from the intrinsic RMs and contributions of the extragalactic path.  The robustness of the JF12 fit to using different masks and subtracting or not subtracting Galactic foreground where known, is a testimonial to the basic approach of de-weighting datapoints with large variance.  See \cite{jf12a} for details.  

JF12 found their best fit when the orientation direction of the striated field is locally aligned with the direction of the coherent field, so that the striated contribution amounts simply to an enhancement of the polarized synchrotron signal relative to the RM signal \cite{jf12a,jf12b}.   This suggests that compression of a coherent field, e.g., by SNe explosions, may be the origin of the striated field effect \cite{fCRAS14}. 

A major source of uncertainty in the GMF comes from our imperfect knowledge of the $n_{\rm e}$ and $n_{\rm cre}$ distributions, on which the observables depend.  The $n_{\rm e, \, cre}$ densities used by JF12 were from NE2001 \cite{NE2001} and GALPROP (A. Strong, private communication), but the GALPROP $n_{\rm cre}$ model is azimuthally symmetric, whereas a more realistic model is likely to have more structure, e.g., \cite{benyamin+13}.  Fortunately, modeling the Galactic CR distribution is a topic of intense effort and more realistic models are beginning to emerge.  The qualitative impact of a more structured $n_{\rm cre}$ is to reduce the strength of the JF12 field in the plane, as discussed in \cite{fCRAS14}.  

\begin{figure}[b!]
\centering
\includegraphics [width=0.85\columnwidth, angle=-90]{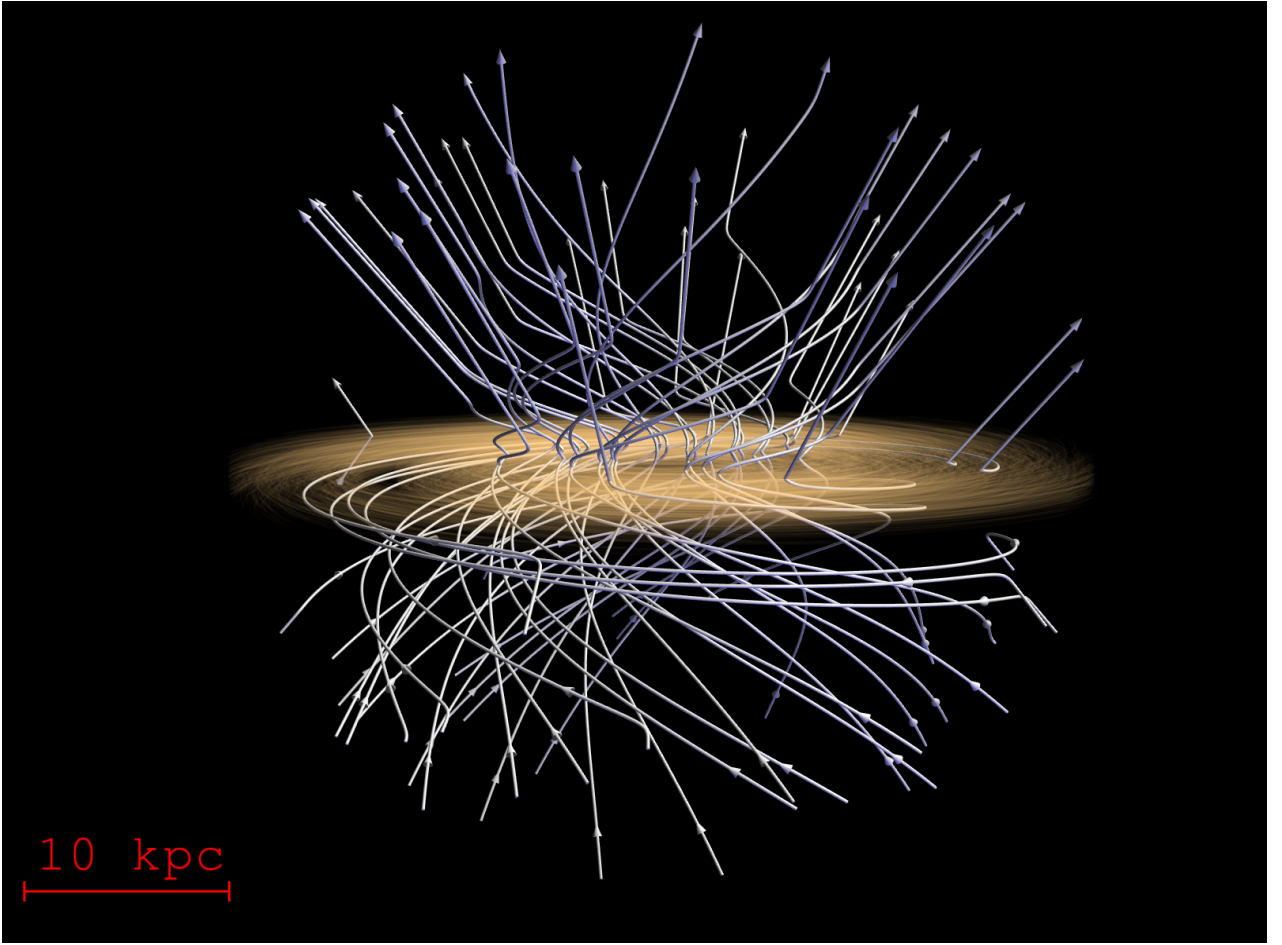}
\vspace{-1pc}
\caption{The outwardly-spiraling halo field of the Galaxy (JF12); the field lines run from S to N.   Visualization produced by T. Sandstrom, NASA. }
\vspace{+2pc}
\label{GMF}
\end{figure}

\begin{figure}
\vspace{-0.25in}
\centering
\includegraphics[width=0.95 \linewidth]{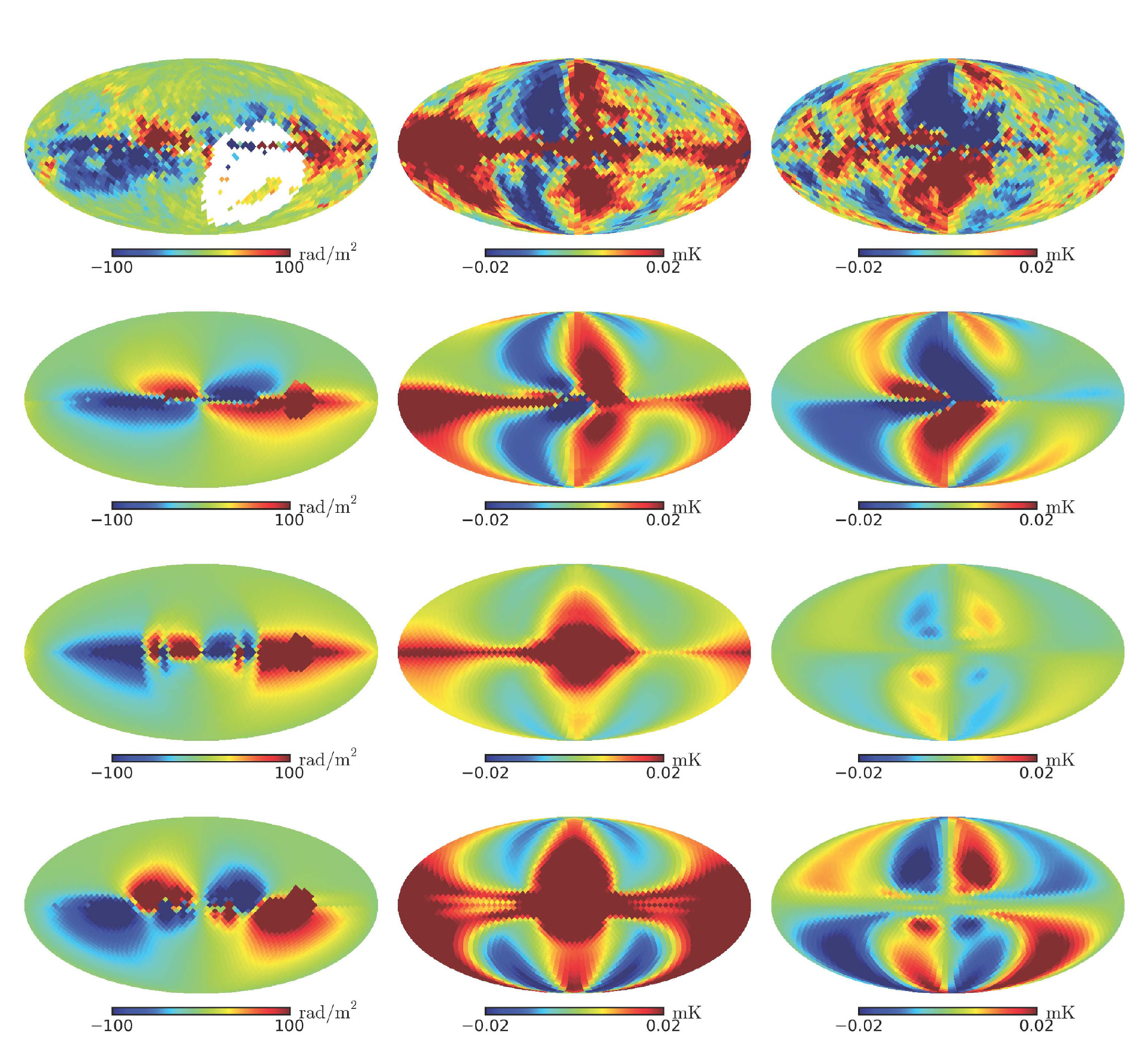}
\vspace{-0.05in}
\caption{Skymaps in Mollweide projection of (left to right) RM (in rad/${\rm m}^2$), Stokes $Q$ and Stokes $U$ (in mK);  Galactic longitude is $l=0^\circ$ in the center and increases to the left.  From top to bottom: row 1, data (white pixels (RM) lack data.); rows 2-4 simulated data from the JF12, SR10 and PTK11 models, respectively.  The $Q,\,U$ data were not fit in a masked region where foreground contamination is greatest, see  yet the JF12 fit is quite good in all regions; see \cite{jf12a} for the masks used. }
\label{JF12datafit}
\end{figure}

In spite of inadequacies in current models of $n_{\rm e, \, cre}$, the observational data on RM, Q, U unambiguously demand a large scale opening-spiral component to the halo field with field lines running from South to North and the toroidal component having opposite signs in the N and S hemispheres.  The relative signs of the poloidal and two toroidal components are consistent with what would be produced by differential shear in the Galaxy, starting with the poloidal component.  Fig. \ref{GMF} shows the field lines of the coherent halo field in the JF12 model; the field of the disk is rendered with finer but more dense lines in a different color and without directional arrows, to aid visual clarity.   Fig. \ref{JF12datafit} shows the RM, Q and U data (top line) and the predictions from the JF12 coherent+striated field model (2nd line; n.b.,the random field does not contribute to these observables).  For comparison, the predictions of the Sun and Reich (SR10) \cite{sr10} and Pshirkov, Tinyakov and Kronberg (PTK11) \cite{pshirkov+11} models are also shown.  The characteristic large scale L-R, N-S symmetry and antisymmetry pattern seen in the Q,U data calls for an outwardly-spiraling halo field as is present in the JF12 model but not in the others.  The Q,U data alone could be fit with a purely striated halo field, but the RM data demands that the halo-field has a coherent component of comparable magnitude to the striated component and determines its direction.  Since  $n_{\rm e}$ and $ n_{\rm cre}$ are positive definite, modifying them affects the deduced strength rather than sign of the fields and cannot produce the pattern of the signs of RM, Q, U (barring an unnatural conspiracy between contributions from distant portions of the line of sight where the sign of the field can change). 

\section{CR deflections in the GMF}

Using the NASA supercomputer Pleiades, and the Runge-Kutta CR propagation code CRT \cite{crt10}, we have performed all-sky, fine-grained ($1.26 \times 10^{7}$ pixels) UHECR backracking through the GMF, for rigitides of $10^{\rm lgR}$ V, with lgR = 18.0, 18.1, ... 19.0, 19.2, 19.4, 19.5, 19.6,19.8 and 20.  Inverting these maps, gives us forward-tracking from any specified extragalactic direction.  We have also performed coarser-grained simulations (typically 10-40k pixels) for lower rigidities and recording the CR position every 30 yr, for a local source and a source at the Galactic Center, at rigidities lgR = 16.5, 17, 17.5, 18 and 18.5 to study the anisotropy of transient Galactic sources.   Except otherwise noted, the plots below show results for the JF12 coherent field plus a specific realization of the random field obtained by weighting a Kolmogorov random field (KRF) having coherence length 100 pc, with the JF12 model for the spatial dependence of $B_{\rm rand}^{2}$ \cite{kfs14}.   Some studies with other realizations of the KRF and with other coherence lengths have been performed  \cite{kfs14,fs15} and comparative plots are shown below as space permits.  

\subsection{UHECR deflections}

\begin{figure}
\vspace{-2.8in}
\centering
\includegraphics[width=\textwidth]{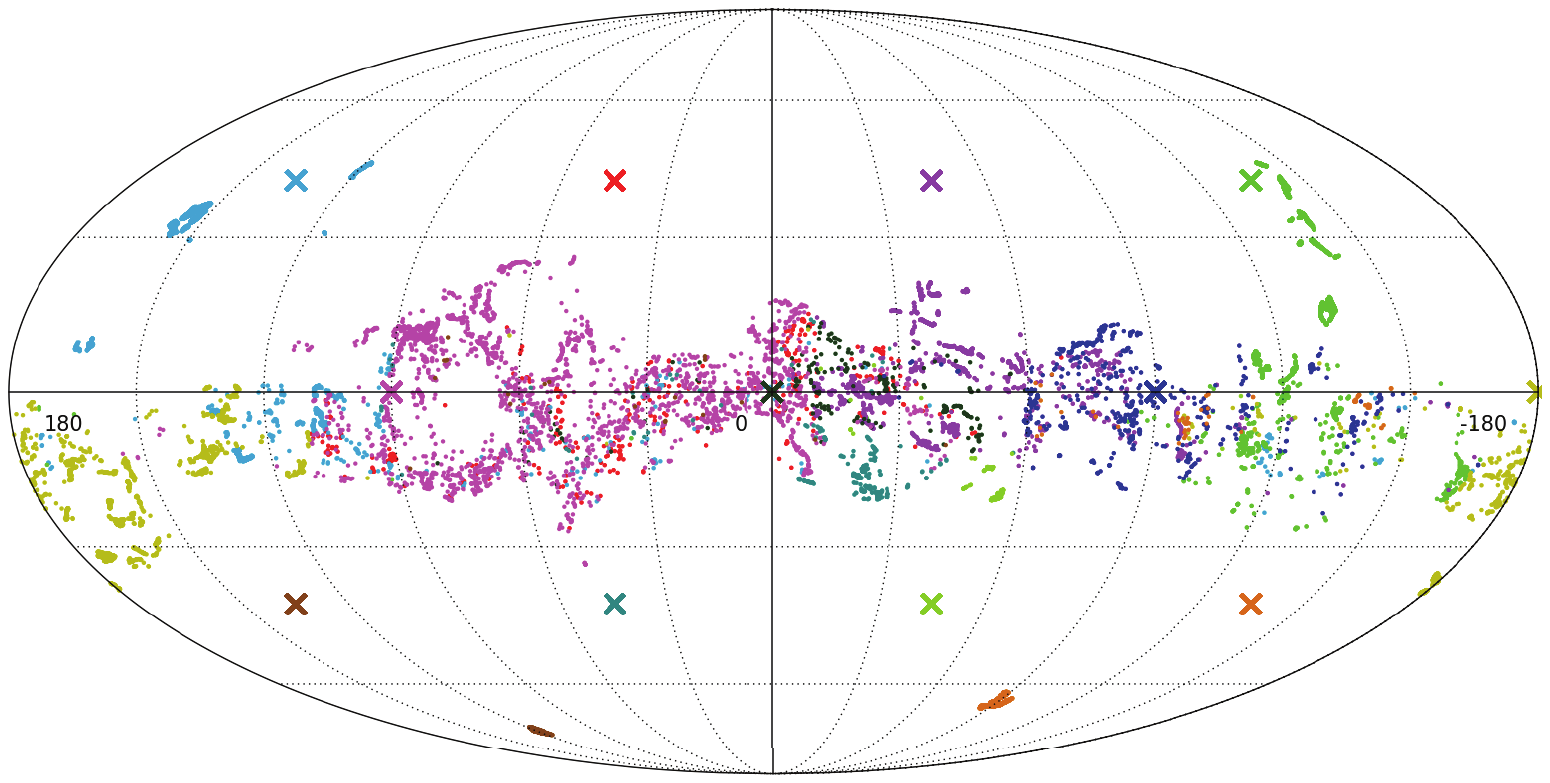}
\vspace{-3in}
\caption{Arrival directions of 10 EV CRs, for 12 representative source directions marked with crosses, at b = $\pm 41.5^{\circ}$, l = $\pm 45^{\circ}$ and $\pm 135^{\circ}$, and at b=0, l = 0, $\pm 90^{\circ}$ and $180^{\circ}$; events from a given source have the same color as the source marker.}  

\label{magdefs}
\end{figure}

Fig.  \ref{magdefs} shows the observed arrival directions for 12 selected source directions -- the centers of the 12 base-level Healpix pixels --  at 10 EV.  One sees that deflections are generally very large.  Except for the 4 directions farthest from the GC (b = $\pm 41.5^{\circ}$, l = $ \pm 135^{\circ}$, the arrival directions are typically highly dispersed.  The Galactic plane is something of an ``attractor'', apart from the directions in the south and away from the galactic center, for which the deflections cause the arrival directions to come from further to the S. By comparing different realizations of the random field, one sees -- not surprisingly -- that the average deflection is basically the same as it depends mainly on the coherent field, while the degree of dispersion decreases when the coherence length or $B_{\rm rand}$ is reduced.  Space does not permit inclusion of more examples, but as would be expected, mean deflections generally increase with decreasing rigidity.  At higher rigidity, such that the deflections do not carry the UHECR too close to the  Galactic plane, the deflections can be much smaller, especially for directions away from the Galactic center.

The GMF acts as a lens for UHECRs \cite{hmrLensing00}.  Although the total number of UHECRs is conserved, and an isotropic distribution of sources leads to an isotropic observed sky, the total flux from any given source direction can be magnified or demagnified relative to the flux in the absence of the GMF, because smaller or larger areas of the plane wave of UHECRs coming from a given source direction can be focused onto the Earth (for illustrative figures see \cite{fCRAS14}).  In the case that multiple distinct regions of the source plane are focussed onto Earth, multiple images of the source are produced, as can be seen in Fig. \ref{magdefs} and discussed in \cite{kfs14} for Cen A as the source.  

\begin{figure}[t]
\centering
\includegraphics [width=0.85\columnwidth]{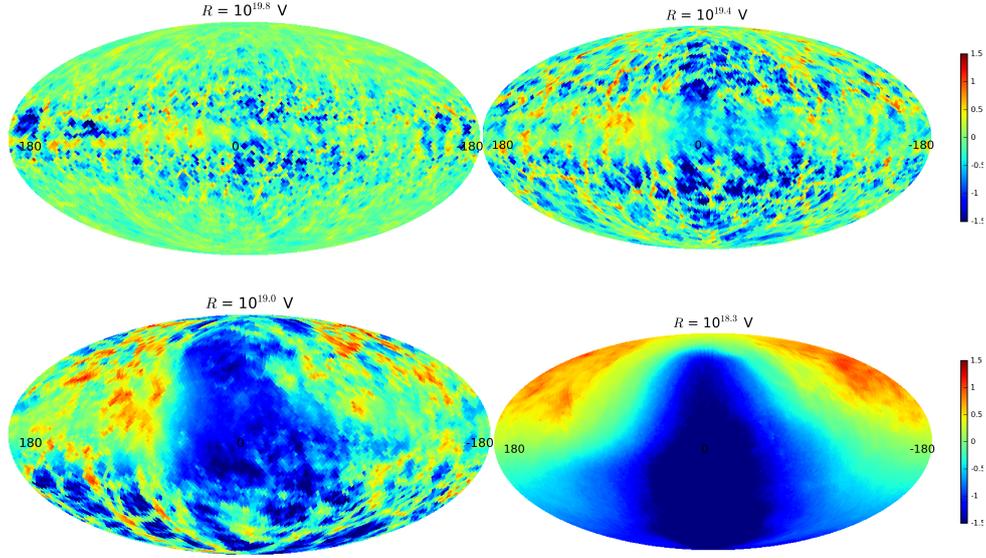}
\vspace{-3pc}
\caption{Skymaps showing log$_{10}$ of the magnification, for 4 different rigidities.  Blue means no CRs from that source direction reach Earth (they are deflected away from us by the GMF) while a source direction indicated with the deepest red has a factor 30 magnification.  For some source directions there is a rapid shift from magnification to demagnification as the rigidity changes.}
\label{magmap}
\end{figure}

While the existence of magnetic lensing is to be expected, the actual impact of the Milky Way magnetic field on UHECR arrival directions can only now be explored, thanks to the new high-resolution simulations and use of a realistic field model.  The results are quite remarkable.  Fig. \ref{magmap} shows skymaps of log$_{10}$ of the magnification at various rigidities, as a function of source direction on the sky.  Dark blue represents complete blindness to the given direction -- UHECRs from those directions are deflected away by the GMF and totally miss Earth -- while the darkest red represents a magnification by a factor 30.   
As the random field is changed, e.g., reduced in strength or the coherence length changed, the details of the pattern of magnification changes, but qualitatively the picture is the same.  The most important features of the magnification map are: 
\begin{itemize}
\item  At high rigidity, the angular size of high-magnification and blind regions is small and their position varies with random field model.
\item For a given random field configuration, the magnification and demagnification for a given source direction can vary rapidly with rigidity.
\item As the rigidity decreases, the blind region grows until about half the sky is invisible --  from behind the Galactic center and especially from the South.
\end{itemize}

\begin{figure}[t]
\centering
\includegraphics [width=0.45\columnwidth, angle=-90]{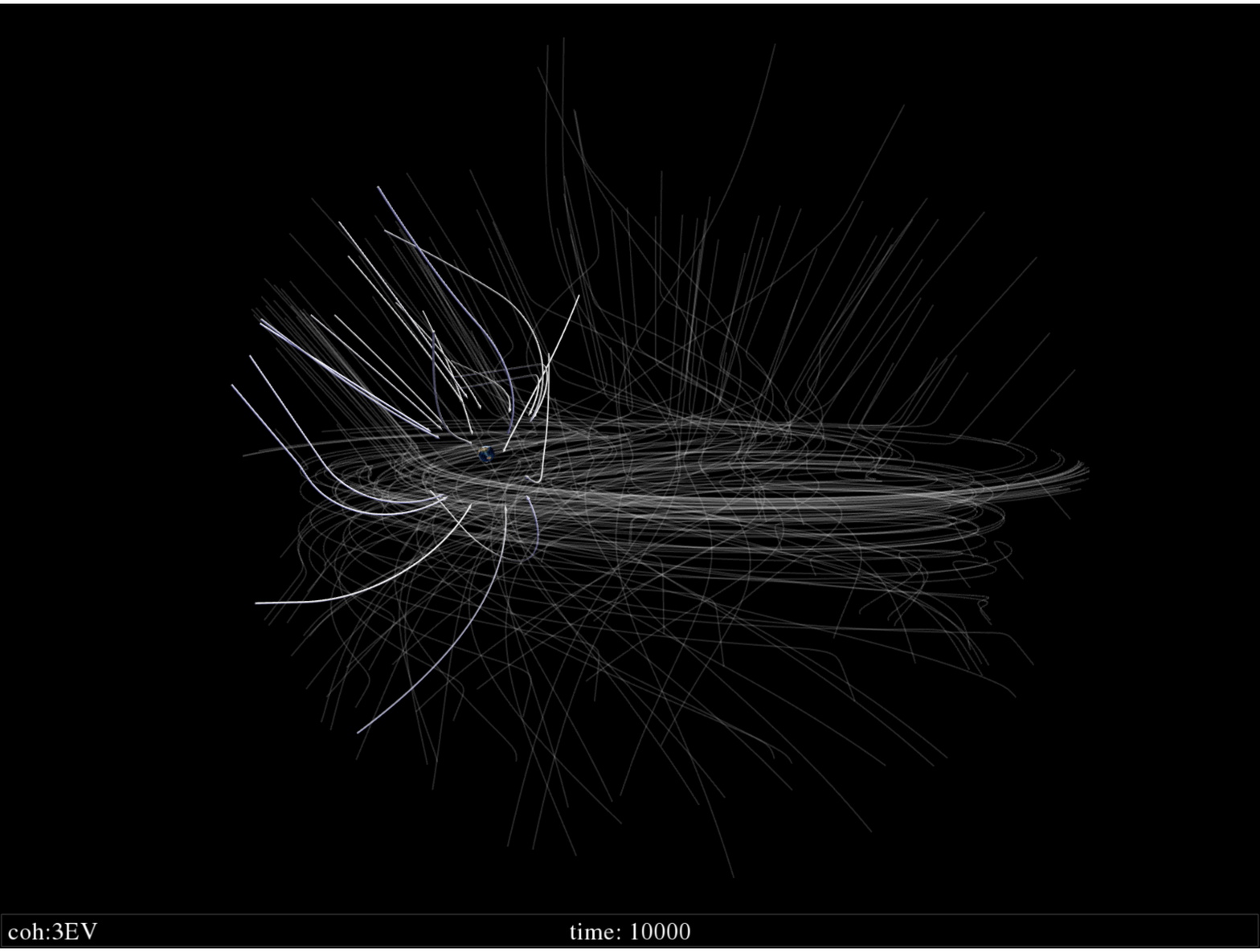}
\includegraphics [width=0.45\columnwidth, angle=-90]{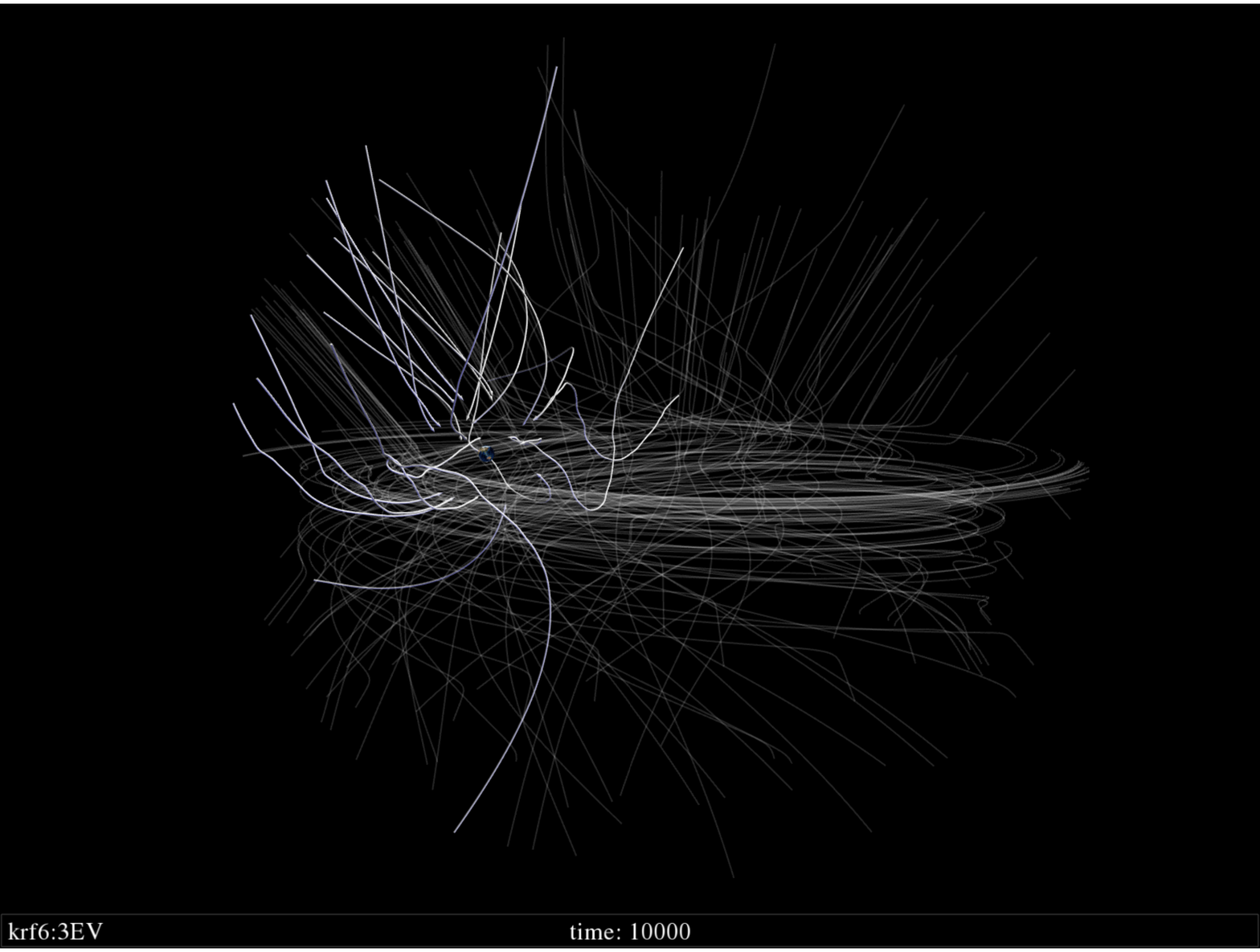}
\vspace{-0.5pc}
\caption{The tracks of 20 UHECRs chosen at random from an isotropically-arriving distribution, for rigidity = E/Ze = 3 EV; the left panel is with the coherent field only, and the right with a 100 pc random field.  The magnification of sources away from and N of the GC, and blindness to sources behind the GC and especially to the south, is clearly seen.   
}
\label{EGCRtracks}
\end{figure}

Fig. \ref{EGCRtracks} (left) shows the trajectories of 20 randomly selected events of rigidity = E/Ze = 3 EV arriving to Earth through the coherent field only, or (right) with random field also, shortly before their arrival.    

\subsection{VHE Galactic CR explosions}

\begin{figure}[t]
\centering
\vspace{-1.5 pc}
\includegraphics [width=0.85\columnwidth, angle=-90]{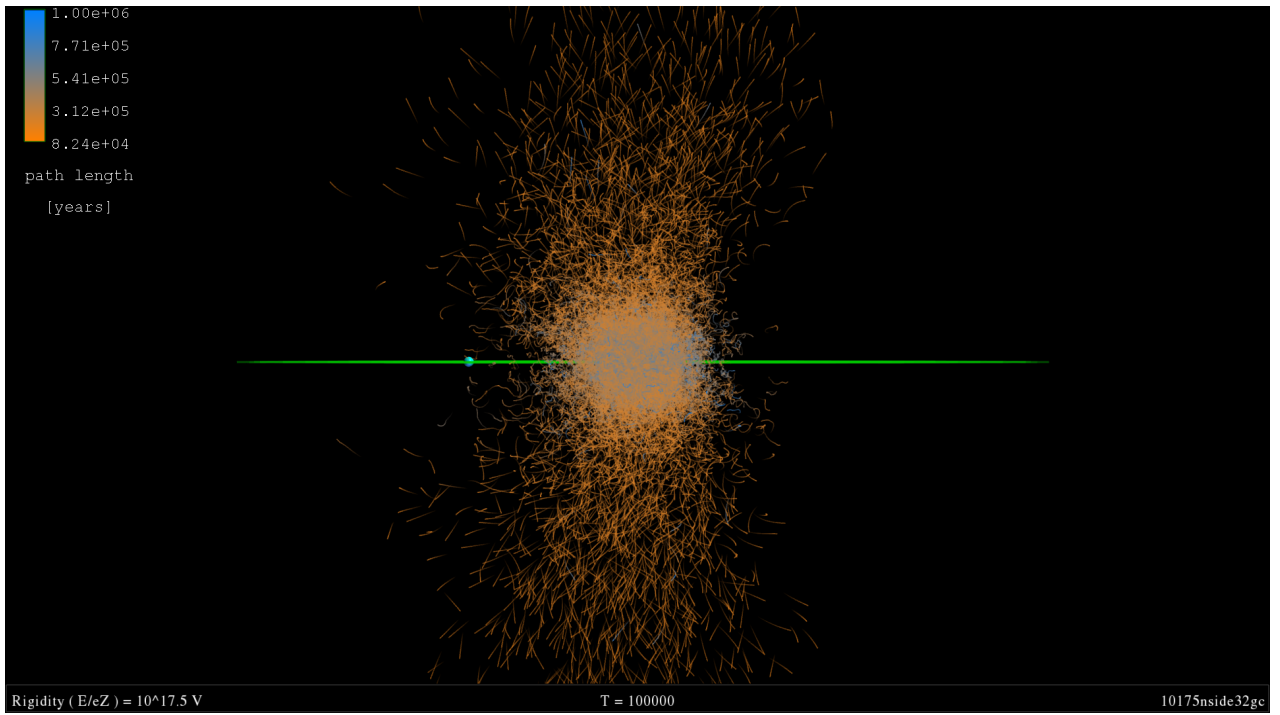}
\vspace{-3.5pc}
\caption{The tracks of high energy (rigidity = E/Z = 0.3 EV) Galactic Cosmic Rays produced at the Galactic Center.  Note the strong anisotropy in diffusion -- almost no cosmic rays reach Earth before escaping the Galaxy.  
}

\label{GCsource}
\end{figure}

Fig. \ref{GCsource} shows a snapshot of 0.3 EV CRs from an explosion at the Galactic Center, 100 kyr after the explosion.  The important point to notice is how anisotropic the ``diffusion'' is -- preferentially escaping vertically.  Almost no CRs reach the solar circle at these energies.   These results underline the crucial importance of including anisotropic diffusion in a realistic magnetic field, before drawing conclusions about galactic cosmic rays, at any energy.

\section*{Acknowledgments}
This research was supported in part by the U.S. National Science Foundation (NSF), Grant
PHY-1212538 and the James Simons Foundation. 


\begin{thebibliography}{10}

\bibitem{jf12a}
R.~{Jansson} and G.~R. {Farrar}, {\it {A New Model of the Galactic Magnetic
  Field}},  {\em ApJ} {\bf 757} (2012) 14.

\bibitem{jf12b}
R.~{Jansson} and G.~R. {Farrar}, {\it {The Galactic Magnetic Field}},  {\em
  Astrophys. J.} {\bf 761} (2012) L11.

\bibitem{fCRAS14}
G.~R. {Farrar}, {\it {The Galactic magnetic field and ultrahigh-energy cosmic
  ray deflections}},  {\em Comptes Rendus Physique} {\bf 15} (Apr., 2014)
  339--348, [\href{http://arxiv.org/abs/1405.3680}{{\tt arXiv:1405.3680}}].

\bibitem{NE2001}
J.~M. {Cordes} and T.~J.~W. {Lazio}, {\it {NE2001.I. A New Model for the
  Galactic Distribution of Free Electrons and its Fluctuations}},
  \href{http://arxiv.org/abs/astro-ph/0207156}{{\tt astro-ph/0207156}}.

\bibitem{benyamin+13}
D.~{Benyamin}, E.~{Nakar}, T.~{Piran}, and N.~J. {Shaviv}, {\it {Recovering the
  Observed B/C Ratio in a Dynamic Spiral-armed Cosmic Ray Model}},  {\em \apj}
  {\bf 782} (Feb., 2014) 34, [\href{http://arxiv.org/abs/1308.1727}{{\tt
  arXiv:1308.1727}}].

\bibitem{sr10}
X.-H. {Sun} and W.~{Reich}, {\it {The Galactic halo magnetic field revisited}},
   {\em Research in Astronomy and Astrophysics} {\bf 10} (Dec., 2010)
  1287--1297, [\href{http://arxiv.org/abs/1010.4394}{{\tt arXiv:1010.4394}}].

\bibitem{pshirkov+11}
M.~S. {Pshirkov}, P.~G. {Tinyakov}, P.~P. {Kronberg}, and K.~J. {Newton-McGee},
  {\it {Deriving the Global Structure of the Galactic Magnetic Field from
  Faraday Rotation Measures of Extragalactic Sources}},  {\em \apj} {\bf 738}
  (Sept., 2011) 192, [\href{http://arxiv.org/abs/1103.0814}{{\tt
  arXiv:1103.0814}}].

\bibitem{crt10}
M.~Sutherland, B.~Baughman, and J.~Beatty, {\it {CRT}: A numerical tool for
  propagating ultra-high energy cosmic rays through {Galactic} magnetic field
  models},  {\em Astroparticle Physics} {\bf 34} (Nov., 2010) 198--204.

\bibitem{kfs14}
A.~{Keivani}, G.~R. {Farrar}, and M.~{Sutherland}, {\it {Magnetic deflections
  of ultra-high energy cosmic rays from Centaurus A}},  {\em Astroparticle
  Physics} {\bf 61} (Feb., 2015) 47--55,
  [\href{http://arxiv.org/abs/1406.5249}{{\tt arXiv:1406.5249}}].

\bibitem{fs15}
G.~R. {Farrar} and M.~{Sutherland}, {\it Deflections of ultrahigh energy cosmic
  rays in a realistic model of the galactic magnetic field},  {\em in
  preparation} (2015).

\bibitem{hmrLensing00}
D.~{Harari}, S.~{Mollerach}, and E.~{Roulet}, {\it {Signatures of galactic
  magnetic lensing upon ultra high energy cosmic rays}},  {\em Journal of High
  Energy Physics} {\bf 2} (Feb., 2000) 35,
  [\href{http://arxiv.org/abs/astro-ph/0001084}{{\tt astro-ph/0001084}}].

\end{thebibliography}

\providecommand{\href}[2]{#2}\begingroup\raggedright\endgroup

\end{document}